\documentclass[fleqn,10pt,twocolumn]{wlscirep}
\usepackage[utf8]{inputenc}
\usepackage[T1]{fontenc}
\usepackage{bm}
\title{Graphite forms via annihilation of screw dislocations}
\author[1,$\dagger$]{Jacob W. Martin}
\author[1,{$\dagger$}]{Jason L. Fogg}
\author[1]{Kate J. Putman}
\author[1]{Gabriel Francas}
\author[1]{Ethan P. Turner}
\author[1]{Nigel A. Marks}
\author[1,*]{Irene Suarez-Martinez}

\affil[1]{Department of Physics and Astronomy, Curtin University, Perth 6845, Australia}

\affil[*]{e-mail: I.Suarez-Martinez@curtin.edu.au}
\affil[$\dagger$]{Joint first authorship}

\begin{abstract}
\Large
Graphite is the thermodynamically stable form of carbon, and yet is remarkably difficult to synthesise.
A key step in graphite formation is the removal of defects at high temperature ($>$2300~$^{\circ}$C) that allow graphenic fragments to rearrange into ordered crystallites~\cite{Oberlin1984}. 
We find the critical defect controlling graphitisation is a screw dislocation that winds through the layers like a spiral staircase, inhibiting lateral growth of the graphenic crystallites ($L_a$) and preventing AB stacking of Bernal graphite.
High-resolution transmission electron microscopy (HRTEM) identifies screws as interdigitated fringes with narrow focal depth in graphitising polyvinyl chloride (PVC).
Molecular dynamics simulations of parallel graphenic fragments confirm that screws spontaneously form during heating, with higher annealing temperature driving screw annihilation and crystallite growth.
The time evolution of graphitisation is tracked via X-ray diffraction (XRD), showing the growth of $L_a$ and reduction of the interlayer spacing consistent with molecular dynamics of screw annihilation.
This mechanistic insight raises opportunities to lower the barrier for graphitisation as well as broadening the range of carbonaceous materials that can turn into graphite, thereby lowering the cost of synthetic graphite used in lithium-ion batteries, carbon fibre, and electrodes for smelting. 
\end{abstract}

\begin{document}

\flushbottom
\maketitle

\thispagestyle{empty}

\section*{Main text}
\begin{figure*}[t!]
\centering
\includegraphics[width=\linewidth]{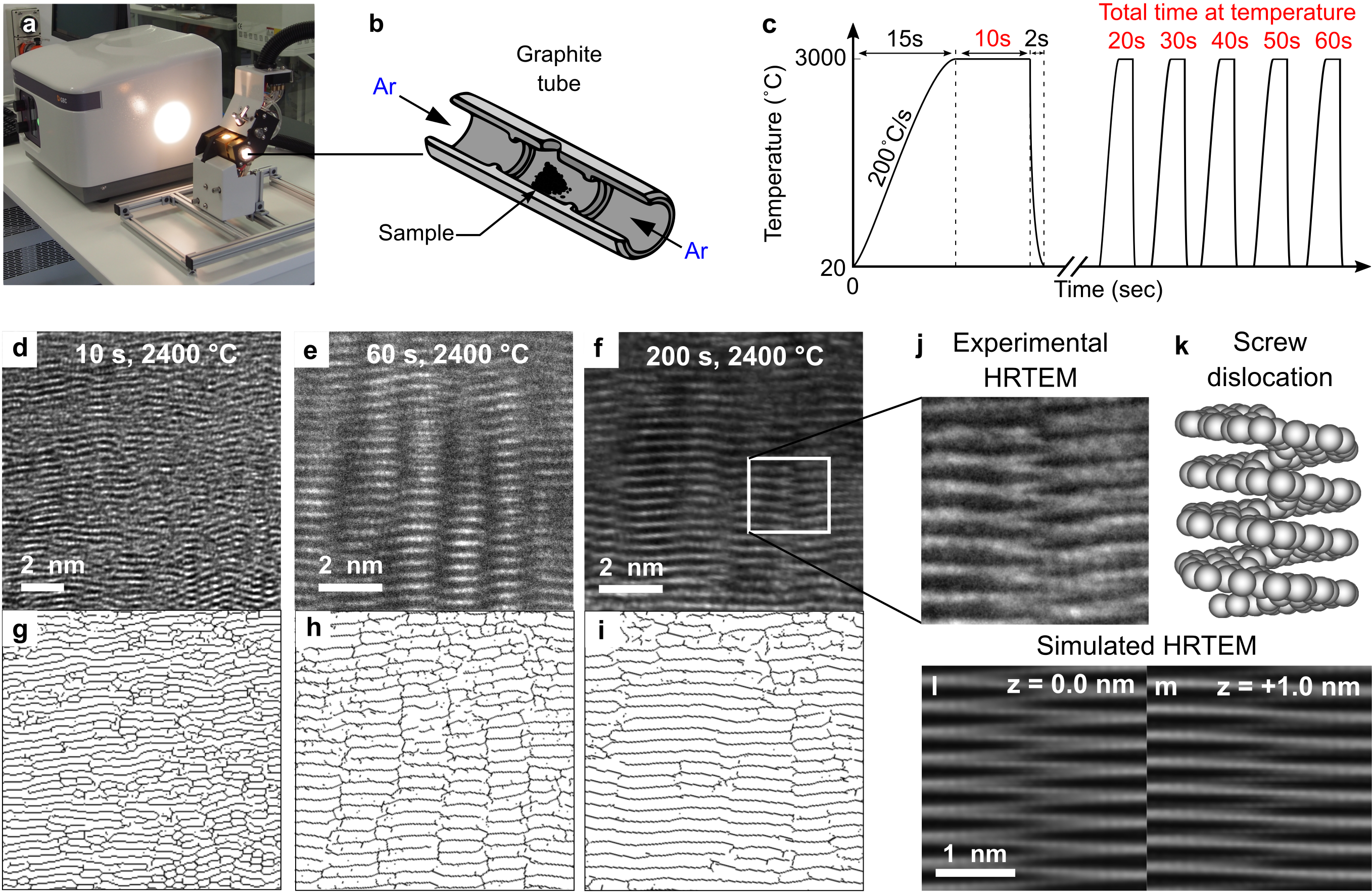}
\caption{\textbf{Observing screw dislocation in partially graphitised PVC}. a) Operational pulsed graphite furnace showing the blackbody radiation from the graphite tube being heated to 2500~$^\circ$C. b) Crossection of the graphite tube with sample that sits within the furnace workhead. c) Repeated pulse sequenced used to probe time dependence of PVC graphitisation. d-f) phase contrast transmission electron microscopy images and g-i) are image processed skeletonised fringes. d) After a 10 seconds pulse at 2400~$^{\circ}$C\ a zig-zag texture of small fringes is found. Many interplanar defects appear. e) For 60 seconds of pulsing the interplanar defects have aligned into columns. f) These interplanar defects begin to annihilate forming extended fringes after 200~s. j) Expanded view of a single interdigitated fringe revealing a zig-zag ramped structure. k) An atomisitic model of a screw dislocation in graphite. l-m) Simulated HRTEM for a screw in AA graphite at focus and a defocus of 1~nm.}
\label{fig:HRTEM}
\end{figure*}

Polyvinyl chloride (PVC) is heated in a custom pulsed graphite furnace\cite{Putman2018,Fogg2020} that reaches up to 3000~$^{\circ}$C in seconds (Fig.~\ref{fig:HRTEM}a-c). 
Repeated pulsing enables the study of graphitisation as a function of temperature and time.
HRTEM images in Fig.~\ref{fig:HRTEM}d-f show the time evolution of graphitisation at 2400~$^\circ$C. 
Figures~\ref{fig:HRTEM}g-i show the same images processed with a skeletonisation algorithm that highlights the dark fringes corresponding to graphenic sheets  at the Bragg angle.
For the 10~s sample (Fig.~\ref{fig:HRTEM}d,g), small basic structural units (BSUs) are seen where a few layers locally align, but little long-range ordering is observed. 
After 60~s the fringes and interlayer defects align into columns extending over many tens of layers (Fig.~\ref{fig:HRTEM}e,h). 
Skeletonisation reveals neighbouring columns are out of phase in the c-direction, normal to the graphenic layers, by half an interlayer distance forming an interdigitated texture. 
For the 200~s sample (Fig.~\ref{fig:HRTEM}f,i), the columns merge into extended fringes, reducing the number of interlayer defects.
These merged fringes are no longer interdigitated but have regions where they bend forming a wrinkled texture, as first noted by Oberlin~\cite{Oberlin1984}.

\begin{figure*}[t!]
\centering
\includegraphics[width=\linewidth]{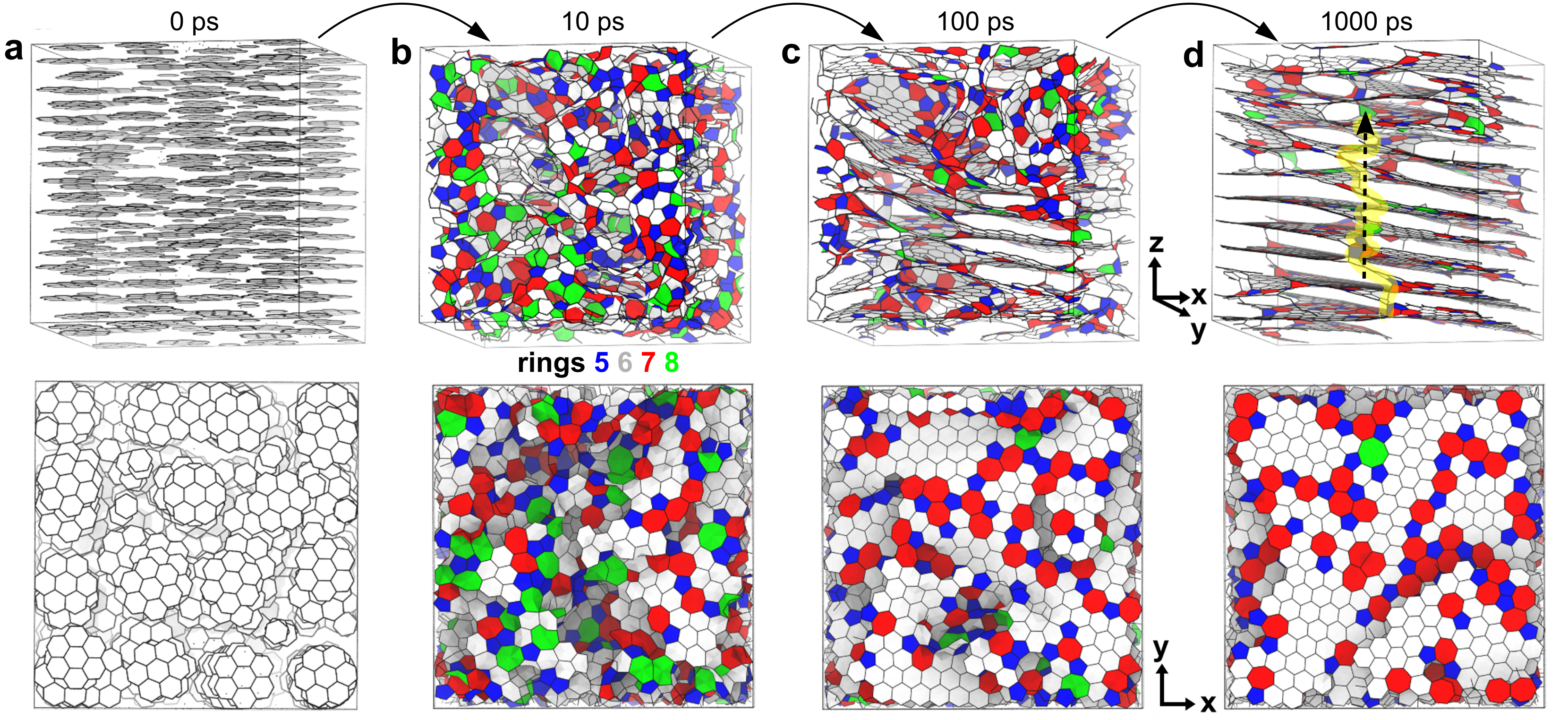}
\caption{\textbf{Screw formation} Top row is a 3D view looking along the basal plane. Bottom row is a plan view looking from above. a) initial configuration of aligned hexagonal regions. b) to d) snapshots of the system during molecular dynamics simulations at 3200~K. Non-hexagonal rings are coloured according to ring number. c) The graphenic regions of hexagonal rings extend as the structure is annealed and screws form between the layers  followed by alignment. d) A long screw is highlighted in yellow.
}
\label{fig:formation}
\end{figure*}

\begin{figure*}[t!]
\centering
\includegraphics[width=\linewidth]{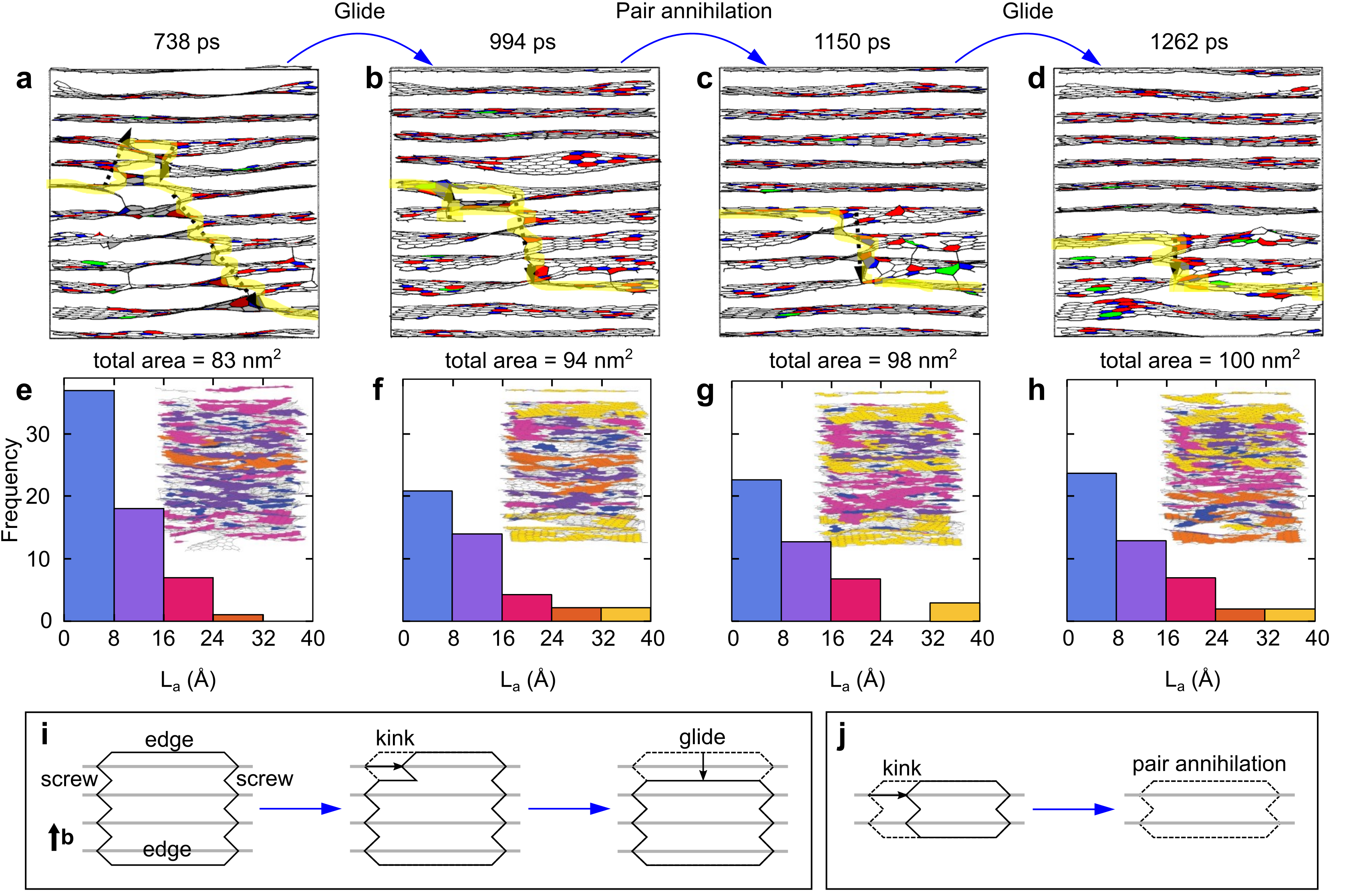}
\caption{\textbf{Screw annihilation} 
a) to d) shows four snapshots of slices of the structure during molecular dynamics simulations performed at 3500~K. Between a) and b) a glide is shown where two screws shrink by a layer. Then from b) to d) two screw dislocations are found to annihilate with the prismatic edge dislocations and further glide leaving the upper layers without interplanar defects. e) to h) shows the total area and histograms for the graphenic crystallite size with inset structures coloured according to their crystallite sizes.
i) shows a schematic of the edge glide mechanism for a dislocation loop. j) shows the pair annihilation of the dislocation loop.
}
\label{fig:annihilation}
\end{figure*}

\begin{figure*}[t!]
\centering
\includegraphics[width=\linewidth]{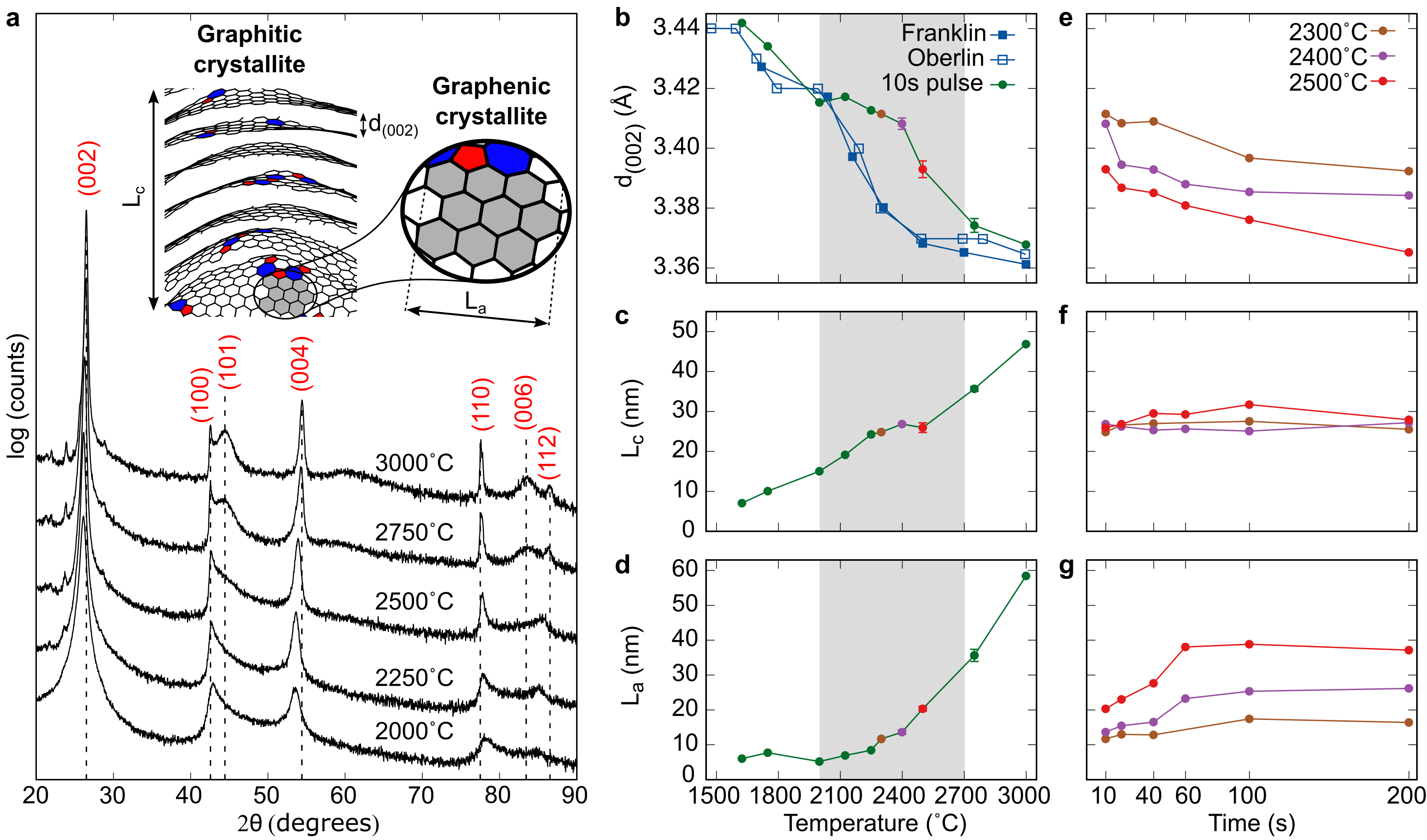}
\caption{\textbf{Observing graphitisation with X-ray diffraction}. a) The temperature dependence on the X-ray pattern for PVC after a single 10~s pulse. The inset shows the interpretation of the (002) reflection as due to the stacking order of a graphitic crystallite. The scattering angle provides the interplanar distance $d_{(002)}$ and the width of the reflection provides the length, $L_{c}$, of the graphitic crystallite in the stacking direction. The (110) reflection arises from the aligned hexagonal regions of a graphenic crystallite where the width of the reflection gives the length, $L_{a\{110\}}$, of the crystallite. b) $d_{(002)}$ is shown for PVC from Franklin \cite{Franklin1951a} and Oberlin~\cite{Oberlin1984} and our 1 pulse 10 second pulsed furnace experiment. The critical temperature is seen to be shifted indicating a non-equilibrium condition. c) $L_{c}$ is seen to continuously transform while d) $L_{a}$ is seen to increase as the $d_{(002)}$ decreases. e) to f) The time evolution of the parameters for 2400, 2500 and 2750~$^{\circ}$C. e) and g), $d_{(002)}$ and $L_{a}$ are seen to change towards equilibrium while $L_{c}$ does not, within experimental uncertainty.}
\label{fig:XRD}
\end{figure*}

We propose that the interdigitated fringes are produced by screw dislocations that connect layers in a helical manner. 
Figure~\ref{fig:HRTEM}j shows a close-up view of the interdigitated fringes, revealing features that zig-zag between the layers.  
The zig-zag feature connects the fringes with ramps that can be seen in an atomistic model of the dislocation core (Fig.~\ref{fig:HRTEM}k). 
To confirm the proposal we compute HRTEM images of screw dislocations at different focal depths with matching instrumental aberrations.
Multislice HRTEM simulations of a single screw in AA graphite at focus (Fig.~\ref{fig:HRTEM}l) reproduces the zig-zag ramp texture, while changing the defocus by 1~nm gives a bent wrinkled texture (Fig.~\ref{fig:HRTEM}m).
This short focal depth is consistent with our experimental observations that changing the defocus by a few nanometres leads to the appearance and disappearance of zig-zags. 
We also ruled out beam damage as a possible source of the interdigitated fringes. The beam intensities used during imaging did not lead to any time evolution in the fringes. Additionally, further increasing of the beam intensity led to amorphisation of the samples rather than promoting graphitisation. 

Our ability to assign the screw dislocation to the interdigitated fringes is enabled by two breakthroughs. 
Firstly, the custom pulsed graphite furnace~\cite{Putman2018} allows for the graphitisation process to be paused.
Repeated pulsing allows us to step through graphitisation and probe intermediate structures, isolating single interdigitated fringes for imaging.

Secondly, the use of a cold field emission gun for the TEM enables higher resolution images by removing the thermal broadening of the electron's beam energy.
The higher resolution reveals ramps connecting the interdigitated layers that have not previously been imaged.
With the benefit of hindsight, both interdigitated fringes and screws are present in the literature, albeit at lower resolution.
For example, TEM by Oberlin~\cite{Oberlin1984} and others~\cite{Ghazinejad2017,Tulic2019,Mubari2022} show textures and general features matching our observations. 
Screws are a well known defect in graphite and can be observed on the surface of graphite using optical microscopy, electron microscopy and atomic force microscopy~\cite{Hennig1965,Rakovan2002}. 
Yakobson \textit{et al.} even assigned interdigitated fringes to screw dislocations to explain the structure of coal~\cite{Sun2011}.
Leyssale \textit{et al.} identified screws in their electron microscopy image guided atomistic reconstruction (IGAR) model  of irradiated graphite~\cite{Leyssale2012}.
However, none of these works assigned the defect in the context of graphitisation, nor did they study the time dependence at a fixed temperature.

To understand how the screws form and annihilate we performed molecular dynamics simulations at high temperature.
Figure~\ref{fig:formation}a shows the initial configuration of graphene fragments (dehydrogenated coronene and benzene) that are parallel to each other but do not form planes.
This arrangement mimics the aligned mesophase formed during carbonisation of graphitising carbons~\cite{Oberlin1984}.
Elevated temperatures of 3200~K were used to enable the dynamics to occur on picosecond timescales.
After 10~ps  (Fig.~\ref{fig:formation}b), the fragments fuse forming many non-hexagonal rings, which are coloured according to ring number. 
After 100~ps (Fig.~\ref{fig:formation}c) small stacked regions become visible and extended hexagonal regions develop. 
By 1000~ps (Fig.~\ref{fig:formation}d) several screw dislocations are formed and the longest screw, highlighted in yellow, runs almost the entire length of the simulation cell.
The core of the screws is dominated by sp$^2$ bonded atoms and present slightly distorted versions of the idealised screw shown in Fig.~\ref{fig:HRTEM}k.
These simulations reveal that screws spontaneously form from an ordered precursor, merging the graphenic regions in the mesophase that are parallel but are not coplanar.

Repeating the MD simulations at a higher temperature of 3500~K shows the annihilation of screw dislocations.
Figure~\ref{fig:annihilation}a-d shows a slice of the system with screw dislocations highlighted in yellow and an arrow in the direction of the dislocation line. 
As the simulation progresses screws are removed by gliding followed by pair annihilation, which increase the number of perfectly stacked layers.
In the upper-left quadrant of Fig.~\ref{fig:annihilation}a a dislocation loop can be seen where two screws are connected by prismatic edge dislocations.
This dislocation loop shrinks through edge gliding that reduces the height of the dislocation loop until it is annihilated.
A schematic in Fig.~\ref{fig:annihilation}i shows the mechanism by which the dislocation shrinks: (i)
a kink forms at the boundary between the screw and prismatic edge, (ii) the kink migrates along the edge dislocation, (iii) the edge reaches the opposing screw, reducing the height by one layer.
In some circumstances kinks can initiate from all four boundaries and meet in the middle.
Once the dislocation loop has shrunk to two layers, it can annihilate in a process which is seen between Fig.~\ref{fig:annihilation}b and c and shown schematically in Fig.~\ref{fig:annihilation}j. 
The annihilation mechanism involves a kink which propagates along basal plane, collapsing the edge dislocations together thereby separating the two layers.
A version of this annihilation mechanism was previously modelled with DFT and referred to as a mezzanine defect~\cite{Trevethan2013,Vuong2017a,Vuong2017b}.

The growth of graphenic crystallites can be tracked in the simulation by considering the hexagonal rings that are connected and closely aligned. Figure~\ref{fig:annihilation}e-h shows a histogram of the crystallite size, L$_a$, with an inset showing the structure colour-coded by fragment size.
At the beginning of the time series a significant fraction of the crystallites are below 24~\AA.
After the top layers are freed from the screw defects the total area of the graphenic crystallites is increased from 83 to 94~nm$^2$ and the regions of extended hexagonal rings grow up to 40~\AA.
This growth is achieved by the removal of non-hexagonal rings through Stone-Wales/Diene transformations~\cite{Dienes1952,Wales1998}.
These screw dislocations appear to pin the in-plane growth of graphenic regions in a similar manner to how dislocations can inhibit growth of grain boundaries in metals.

The growth of crystallites during graphitisation can also be tracked via XRD.
The most important XRD parameter for studying graphitisation is the interlayer distance $d_{(002)}$ which has a minimal value of 3.35\AA\ for graphite.  
XRD also determines the size of the crystallites via the quantities $L_c$ and $L_a$ (see inset in Fig.~\ref{fig:XRD}a).
Figure~\ref{fig:XRD}a shows the XRD pattern for PVC after a single 10~s pulse.
At 3000~$^{\circ}$C the pattern shows all the signatures of highly graphitised material, including a sharp (002) reflection, (101) and (112) reflections signifying 3D ordering, and higher order (004) and (006) peaks. 
At 2000~$^{\circ}$C the 3D peaks are absent and the other peaks are considerably broader, indicating that graphitisation has yet to occur.

Figure~\ref{fig:XRD}b-d shows the temperature dependence of the these parameters during the graphitisation transition and are compared with literature data from Franklin~\cite{Franklin1951a} and Oberlin~\cite{Oberlin1984}. 
These works used conventional furnaces that take hours to heat and cool~\cite{Chung2002}, meaning samples were at thermodynamic equilibrium. 
At 3000~$^\circ$C and below 2000~$^\circ$C our parameters match the Franklin and Oberlin data, indicating that 10~s is sufficient to achieve equilibrium.
The intermediate regime, highlighted by gray shading, corresponds to a temperature range where equilibrium takes longer than 10~s.
The work of Ouzilleau \textit{et al.} \cite{Ouzilleau2016} provides another way to think about this region.
They analysed data from conventional furnaces and showed that the $d_{(002)}$ parameter for all graphitising carbons follow a universal second-order phase transition with a critical temperature of $T_c=2280\pm 50^{\circ}$C.
We find this critical temperature is shifted to higher temperature of $\sim 2500^{\circ}$C for the 10~s pulse but the same profile for d$_{(002)}$ versus temperature is observed. 

The XRD signature of screws is seen in repeated pulsing measurements in the transition region (Fig.~\ref{fig:XRD}e-g).
With increasing time, $d_{(002)}$ decreases, $L_a$ increases, while $L_c$ remains constant.
This observation is in agreement with our assignment of screws, since annihilation increases the basal extent (\textit{i.e.} increasing La) while leaving the number of layers ($L_c$) unchanged.
Single screw dislocations (Burgers vector of one interlayer distance) only form in AA and ABC graphite leading to an interlayer distance at the screw core of 3.5~\AA~\cite{Martinez2007}. 
As the screws are progressively removed the graphenic planes are increasingly able to adopt AB stacking and $d_{(002)}$ decreases toward the ideal graphite value of 3.35~\AA.
Since screws inhibit AB stacking they provide a plausible explanation for so-called turbostratic graphite in which the graphenic planes are in random registration with an interlayer spacing of 3.42~\AA~\cite{Puech2019}.

Screw dislocations could also explain the twisting disorder seen in graphite and magnetic states. Moir\'{e} patterns have been observed using scanning tunnelling microscopy and electron microscopy these features are caused by layers being twisted relative to each other~\cite{Campanera2007,Boi2021}. 
Firstly, screws could pin these twisted layers inhibiting the rotation to ideal AB stacking. 
Secondly, screws provide a twisting force near their core described by the Eshelby twist.
For small twist angles exotic magnetic states have been found in twisted bilayer graphene, with a magic angle even enabling a superconducting state~\cite{Cao2018}. 
These magnetic states have led to the field of twistronics where 2D layered materials can form useful devices by controlling the twist of their layers.
Recently, Moir\'{e} pattern with these magnetic properties have been seen in highly oriented pyrolytic graphite (HOPG)~\cite{Boi2021}.
Control of screw defects within carbonaceous materials could potentially enable control of the twist angle and be used for twistronic applications.

The barrier for screw annihilation could be reduced in graphitising materials via chemical or mechanical approaches.
For example, metals that are found to catalyse graphitisation~\cite{Oya1979} may aid in edge gliding.
One may imagine mechanical shear~\cite{Bonijoly1982}, twisting or stress could lower the barrier for screws to annihilate.  
By targeting screw removal, the temperature of graphitisation could be reduced, lowering the cost of synthetic graphite.

This work also sheds light on why so few materials are graphitisable. 
Ouzilleau \textit{et al.} explored this question and proposed two classes of defects~\cite{Ouzilleau2019}: annealable defects that are removed in all carbon materials, and non-annealable topological defects that inhibit graphitisation in non-graphitising materials. 
Our work suggests what these defects might be.
Non-hexagonal rings, screw and edge dislocations are all annealable topological defects in graphitising carbons. 
All these defects were also found in our previous models of non-graphitising carbon~\cite{Martin2019d}, where we employed the same annealing procedure as in this paper but started from a completely amorphous configuration.
The unique aspect of these non-graphitising carbons is that they possess a foam-like 3D connected structure with net-negative curvature~\cite{Martin2019d}.
This net-negative curvature is enabled by disclinations, which should not be conflated with simply non-hexagonal rings but describe a surface that cannot be flattened (or in the language of topology, non-developable). 
Combining our two models we can assign Ouzilleau's defects: non-hexagonal rings, screw and edge dislocations are annealable and present in all carbons, while disclinations are non-annealable defects and inhibit graphitisation. 
For most carbonaceous precursors, carbonisation leads to crosslinking in three dimensions that results in disclinations. 
This makes early parallel alignment during carbonisation critical for discouraging disclination formation and explains why the most common approach is using a precursor that decomposes into a mesophase with aligned aromatic molecules.
Alignment of the mesophase provides a retrospective explanation for the many approaches (shear in carbon fibres~\cite{Ghazinejad2017}, stress in HOPG~\cite{Blackman1962} and 2D templating~\cite{Nishihara2012}) that have been empirically developed to promote graphitisation. 

In summary, we identify screw dislocations as the key defect in graphite formation. Using computer simulations, we provide an atomistic mechanism for screw formation and annihilation, which is consistent with transmission electron microscopy and X-ray diffraction. Efforts to target this defect will be key to lowering the cost of synthetic graphite and enabling more materials to be converted into graphite. 

\section*{Methods}

\subsection*{Preparation of carbonised and graphitised samples}
Polyvinyl chloride was heated in a STF 1200 tube furnace over four hours to a carbonisation temperature of 1000$^\circ$C. The temperature was held for a further hour before passively cooling to room temperature. Argon, flowing at a constant 1.5-2~L/min maintained an inert atmosphere. The carbonised sample was then milled in a mortar and pestle before further heating.

Temperatures above 1000$^\circ$C were achieved using a customised furnace built by GBC Scientific Equipment. The system separates the joule heated graphite tube of an atomic absorption spectrometer from the spectrometer, leaving a bench top unit capable of heating to 3000$^\circ$C in 1--2 seconds. Elevated temperatures can be maintained for several tens of seconds, whilst an inert cooling gas, typically argon, is followed through the tube to prevent combustion. Operation in a pulsed mode extends time at elevated temperature, a short pause between pulses allows the cooling system to manage the thermal load. The temperature profile for heating pulses is as follows: room temperature~-~100$^\circ$C in 5~seconds, ramping to desired temperature at 200$^\circ$C/s, holding desired temperature for a further 10~seconds before cooling to room temperature over a few seconds. The graphite tube holds samples on the tens of milligram scale.

\subsection*{Characterisation of partially graphitised material}
X-Ray Diffraction (XRD) is performed in a Bruker D8 Advance Diffractometer with Bragg-Brentano geometry and a Cu K-$\alpha$ source. XRD specimens are prepared by placing a powder sample onto a low signal silicon wafer holder. Patterns are collected over a 2$\theta$ range of 10-90$^\circ$ with a dwell time of 1.5~seconds and step size of 0.3$^\circ$. Further information regarding the device configuration can be found in Ref.~\cite{Putman2018}. Transmission electron microscopy samples are prepared by grinding with ethanol in an agate mortar, the suspension is sonicated before deposition on 300 mesh lacey carbon grids. A JEOL F200 TEM fitted with a  cryogenic electron source is used to analyse samples nanostructure. With an accelerating voltage of 200~KeV the cold field emission gun produces a coherent electron beam which allows images to be collected with a low surface current. Specific care is taken to minimise beam damage by limiting sample time under the beam and using low beam intensities. Skeletonisation was performed based on the previously developed methodology from Botero \textit{et al.} but re-implemented in python with skimage with     the thresholding approach of Zack \textit{et al.}~\cite{Zack1977} replacing the Otsu method provided more interplanar features.

\subsection*{Simulations}
Initial atomic configurations were prepared by packing coronene fragments of the same orientation into a 4~nm size cube using the software packmol~\cite{Martinez2009}, benzene fragments were then added until the density reached 2~g/cc. The carbon environment dependent interatomic potential (EDIP) was used to describe the energy of the carbon network~\cite{Marks2000}, which has been shown to accurately describe the bonding in sp\textsuperscript{2} to sp\textsuperscript{3} carbons~\cite{DeTomas2016}. The structures were minimised and molecular dynamics simulations performed using the software LAMMPS ~\cite{Plimpton1995}. The Bussi thermostat maintained the temperature during the simulation~\cite{Bussi2007}. The results were processed in polypy~\cite{Kroes2010} to extract the rings using the Franzblau ring algorithm~\cite{Franzblau1991} and custom scripts extracted graphenic regions. Finally, all of the visualisations were rendered in the software VMD~\cite{Humphrey1996}. Atomsk~\cite{Hirel2015} was used to generate screw dislocations in AA graphite. Multislice software compuTEM~\cite{Kirkland2010} was then used to simulate the HRTEM images of the dislocations using aberrations matching those of the instrument.

\bibliography{references.bib}

\noindent\textbf{Acknowledgements}\\
The authors would like to thank Assoc. Prof. Martin Saunders for his help in collecting the HRTEM images.\\
J.W.M. acknowledges the support of the Forrest Research Foundation.\\
I.S.-M. acknowledges the support of the Australian Research Council (No. FT140100191).\\
Characterisation of the samples were undertaken at the John de Laeter Centre, Curtin University.\\
Computational resources were provided by the Pawsey Centre with funding from the Australian Government and the Government of Western Australia.\\
The authors acknowledge the facilities, and the scientific and technical assistance of Microscopy Australia at the Centre for Microscopy, Characterisation \& Analysis, The University of Western Australia, a facility funded by the University, State and Commonwealth Governments.

\noindent\textbf{Author contributions}\\
J.W.M. and J.L.F. carried out the experimental work. J.W.M., K.J.P., G.F., E.P.T., N.A.M and I.S.-M. carried out the modelling and analysis. All authors discussed and co-wrote the manuscript. \\

\noindent\textbf{Competing interests}\\
The authors declare no competing interests. \\


\setcounter{figure}{0}
        \renewcommand{\thefigure}{E\arabic{figure}}

\end{document}


\flushbottom
\maketitle

\thispagestyle{empty}

\section*{Main text}

Carbonisation and graphitisation has been extensively studied for decades, however, the precise atomistic mechanism remains unknown. In 1965, Brooks and Taylor first observed under polarised light how disc-shaped aromatic molecules within graphitising carbons align during carbonisation, forming a liquid crystal mesophase below 1000~$^{\circ}$C~\cite{Brooks1965}. Further heating leads to extensive crosslinking between these aromatic regions as heteroatoms such as hydrogen, chlorine or oxygen are thermally driven off. X-ray diffraction revealed significant disorder in the structure at this stage with the distance between the layers significantly larger than that of graphite (3.4 versus 3.35~\AA), misalignment twist between the layers away from ideal ABA stacking called turbostratic disorder~\cite{Puech2019} and broad reflections suggesting small regions of stacking and hexagonal ordering. This disorder is only removed between 2000--2500~$^{\circ}$C when graphitisation occurs. 
\begin{figure}[b!]
\centering
\includegraphics[width=\linewidth]{Diagram2.png}
\caption{\textbf{Probing the turbostratic to graphitic phase transition by preparing partially graphitised material.} Carbonised PVC is shown as prepared and in the electron microscope before and after graphitisation. The pulsed graphite furnace is shown that can provide short pulses giving partially graphitised material.}
\label{fig:Schematic}
\end{figure}

In the 1970s, Oberlin and coworkers extensively studied this transformation at the nanoscale using electron microscopy~\cite{Oberlin1984}. Dark field transmission electron microscopy images revealed small regions of aligned stacked layers called basic structural units that initially are misaligned. Further details were found using phase contrast or high resolution imaging where dark and light fringes are formed by small aromatic regions at Bragg angle from the electron beam. Figure~\ref{fig:Schematic} shows these fringes that are misaligned by $\sim$10$^{\circ}$ forming a zig-zag texture. During the transformation fringes align into columnar texture where fringes are interdigited and shifted by half a layer (c/2), then as the fringes enlarge a wavy texture follows before finally the fringes become ordered graphite. In dark field TEM Moir\'{e} fringes can be seen as the graphenic regions extend and align forming the ABA stacked ordering. Most recently, Ouzilleau \textit{et al.} demonstrated that graphitisation follows a second order structural phase transition that occurs with a critical temperature of $T_c=2250^{\circ}$C~\cite{Ouzilleau2016}. These authors suggest stable topological defects that require these extreme temperatures to anneal.

Many defects have been suggested to inhibit graphitisation. Defects involving non-hexagonal rings, specifically 5775 dislocation dipole (otherwise known as a Diene or Stone-Wales defect), were suggested in the work of Ouzilleau \textit{et al.}~\cite{Ouzilleau2016}. Indeed non-hexagonal rings have been directly imaged in disordered carbons~\cite{Harris2008,Guo2012}. These would provide the wavy textures seen and would limit the size of the graphenic regions. Isolated pentagonal rings provide positive disclinations (fullerene-like or bowl-shaped structures) that have been suggested to inhibit graphitisation~\cite{Harris2013}. Heptagonal rings provide negative disclinations (schwarzite-like or saddle-shaped structures). These heptagonal rings have been suggested to interconnect graphite layers, forming wormholes that inhibit graphitisation~\cite{Margine2007}. Other interconnects have been suggested such as interstitial atoms pinning the rotation of sheets or isolated layers that restructure into 'ruck-and-tuck'~\cite{Heggie2011}, ramps/messanines~\cite{Trevethan2013} or Y/T junctions~\cite{McHugh2022}. 

Recently, modelling of self-assembled disordered carbon structures using image reconstruction~\cite{Leyssale2012} and our recent annealed molecular dynamics simulations~\cite{Martin2019d} have demonstrated the significant role of screw dislocations, which we have previously studied in ideal graphite~\cite{Suarez-Martinez2007}. Screw defects are well known in graphite with 1952 imaging showing superscrew dislocations across many layers at concentrations of 10\textsuperscript{2--4}~cm\textsuperscript{--2}~\cite{Horn1952} as well as screws with a single layer (elemental screws) later found in 1965 at concentrations of 10\textsuperscript{6--8}~cm\textsuperscript{--2} \cite{Patel1965,Hennig1965}. Recently, they have been directly imaged using atomic force microscopy~\cite{Rakovan2002}. All of these interplanar defects discussed have been shown to provide wavy and offset fringes seen in HRTEM~\cite{Sun2011,Trevethan2013}. However, critically the defect(s) involved in the graphitisation transition have yet to be observed and related to a mechanistic understanding of the process, inhibiting efforts to enhance graphite formation. What is needed is the ability to probe defects during graphitisation.

In order to probe the defects responsible, partially graphitised material needs to be prepared, however, attempts have been hampered by the need to rapidly heat and cool samples within seconds. Graphite furnaces can provide uniform heating but require hours to heat up meaning they can only probe equilibrium state of carbon materials~\cite{Chung2002}. While laser heating have been shown to rapidly heat samples but are unable to uniformly heat a sample to a defined temperature and also suffer from ablation~\cite{Abrahamson2018}.
Recently, we have developed a pulsed graphite furnace that allows us to heat samples to 3000~$^{\circ}$C within seconds and maintain temperature for less than a minute (see Figure~\ref{fig:Schematic})~\cite{Putman2018,Fogg2020}. This provides us the opportunity of preparing uniformly heated \textit{partially graphitised material} while observing the defects involved.

In this letter, a pulsed high-temperature furnace is used to observe the change in defective structures during graphitisation using X-ray diffraction and high resolution transmission electron microscopy. Molecular dynamics simulations and simulated microscopy images identify the critical defect as screw dislocations. The movement and annihilation of screws is also described.

\begin{figure*}[h!]
\centering
\includegraphics[width=\linewidth]{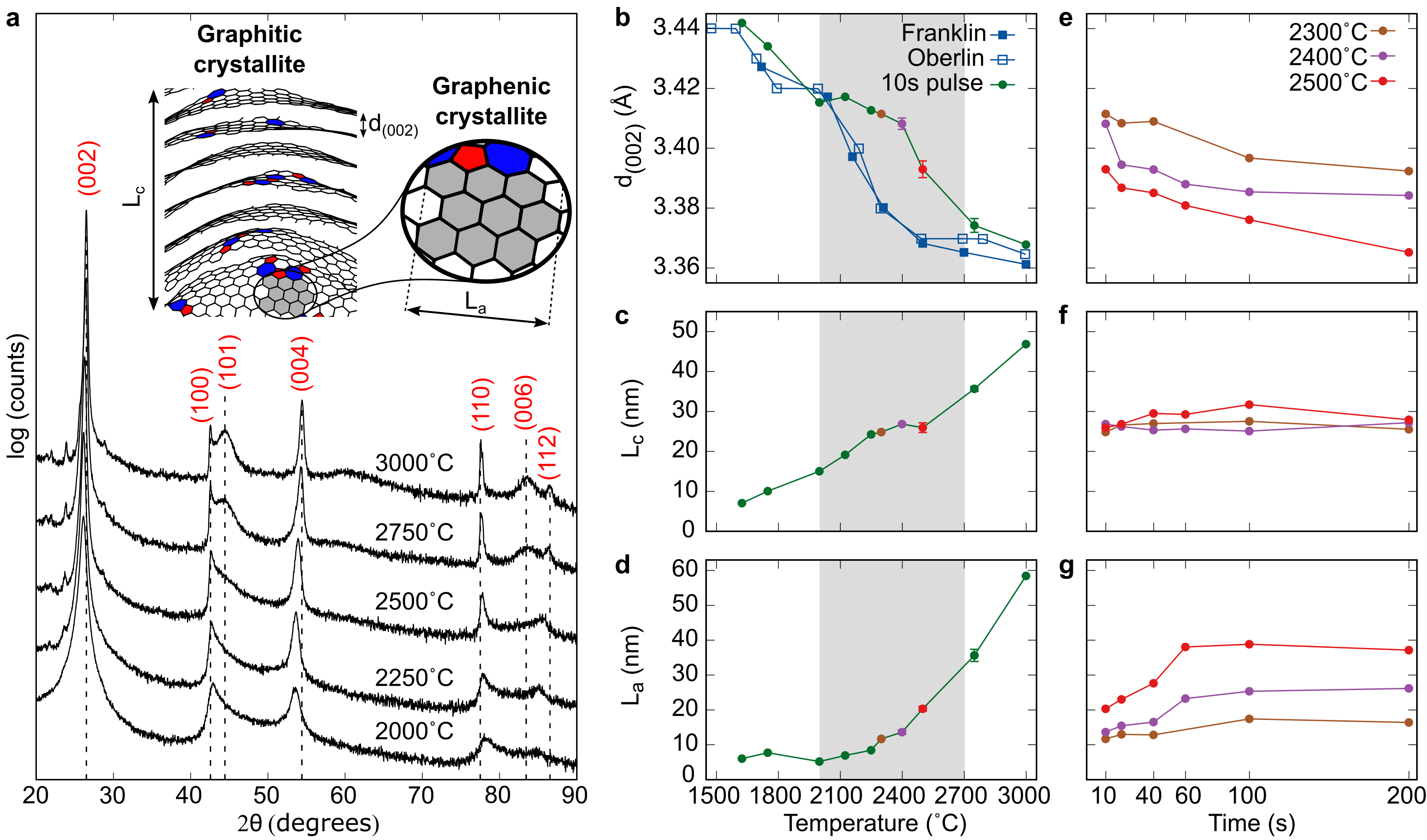}
\caption{\textbf{Observing graphitisation with X-ray diffraction}. a) The temperature dependence on the X-ray pattern for PVC after a single 10~s pulse. The inset shows the interpretation of the \{002\} reflection as due to the stacking order of a graphitic crystallite. The scattering angle provides the interplanar distance $d_{\{002\}}$ and the width of the reflection provides the length, $L_{c\{002\}}$, of the graphitic crystallite in the stacking direction. The \{110\} reflection arises from the aligned hexagonal regions of a graphenic crystallite where the width of the reflection gives the length, $L_{a\{110\}}$, of the crystallite. b) $d_{\{002\}}$ is shown for PVC from Franklin \cite{Franklin1951a} and Oberlin~\cite{Oberlin1984} and our 1 pulse 10 second pulsed furnace experiment. The critical temperature is seen to be shifted indicating a non-equilibrium condition. c) $L_{c\{002\}}$ is seen to continuous transform while d) $L_{a\{110\}}$ is seen to increase after the critical temperature. All of the parameters' time evolution are shown for multiple pulses. For the $d_{\{002\}}$ exponential decays are fitted from a value before the phase transition at 2000~$^{\circ}$C. The $L_{c\{002\}}$ is not seen to change within the error bars which are shown for 2400, 2500 and 2750~$^{\circ}$C.}
\label{fig:XRD}
\end{figure*}

X-ray diffraction provides insights into the kinetics of graphitic (stacking order) and graphenic (hexagonal order) crystallites during graphitisation with the pulsed furnace. 
These structural metrics can be measured from the X-ray diffraction pattern shown in Figure~\ref{fig:XRD} a) (with the inset providing an interpretation of the crystallites with a molecular structure).
The graphitic crystallite comes from the stacking of the layers providing the low angle \{002\} reflection. 
From the peak position the average distance between layers can be determined $d_{\{002\}}$ and the width of the peak provides insight into the size of the regions of stacking $L_{c\{002\}}$.
During graphitisation the $d_{\{002\}}$ is found to change between 2000$^{\circ}$C and 2800$^{\circ}$C approaching the value for graphite of 3.35\AA. The $L_{c\{002\}}$ is also found to continuously change during heating. The graphenic structure comes from small regions where hexagonal rings of carbon atoms in the same plane provides higher angle reflections such as the \{110\} reflection. The peak width then gives an average size of the graphenic crystallites, $L_{a\{110\}}$.

Figure~\ref{fig:XRD} b) shows that after a single 10~s pulse the $d_{\{002\}}$ is shifted to a higher critical temperature of $T_c\sim 2500{\circ}$C. This demonstrates the system after 10~s pulse is out of equilibrium and further pulses will progress the material towards the thermodynamic equilibrium. Figure~\ref{fig:XRD} d) shows that $L_{a\{110\}}$ follows the $d_{\{002\}}$ only increasing after 2400$^{\circ}$C.
However, all three parameters were found to continuously transform during graphitisation after a single pulse at different temperatures.

The pulsed approach also provides a means of probing this phase transition at a fix temperature over time.
For a given set of temperatures multiple pulses can be applied and the system is seen to converge to the equilibrium value of the $d_{\{002\}}$ for that temperature, as does the $L_{a\{110\}}$. However, surprisingly the $L_{c\{002\}}$ is not found to change instead being constant to within the experimental uncertainty. 
This shows that $L_{c\{002\}}$ is in thermodynamic equilibrium and is not involved in the second order phase transition of graphite.
A barrier for graphitisation can be computed using the $d_{\{002\}}$ values from a starting value before graphitisation at 2000~$^{\circ}$C in a similar manner to Ouzilleau \textit{et al.} \cite{Ouzilleau2016}. The barrier was found to be 6$\pm$1~eV (kJ/mol).
Pulsed experiments reveal a non-equilibrium system where $L_{a\{110\}}$  and $d_{\{002\}}$ change dependently while $L_{c\{002\}}$ is found to be independent. 

Further insights can be gleaned from imaging the partially graphitised material at the fixed temperature of 2400~$^{\circ}$C as it equilibrates.
Low electron intensities were used and short exposure times to not disturb the structure and increasing beam intensities were found to increase the amorphous fraction in the material suggesting the ordering is solely due to the high temperature treatment.
Figure~\ref{fig:HRTEM} shows the phase contrast images from the HRTEM as well as skeletonised fringes. With the aid of the cold field emission gun, which enable higher resolution than previously achievable, interlayer defects could also be imaged and seen as vertical skeletonised fringes. 
For a single pulse a series of fringes appear in the misaligned zig-zag texture described earlier~\cite{Oberlin1984}. After 60 seconds an alignment of interplanar defects and fringes is seen. This has been called a columnar structure again by Oberlin and colleagues~\cite{Oberlin1984}.
Finally, after 200 seconds the fringes were found to elongate and the interplanar defects significantly decreased. A wavy texture was found during this phase.
During imaging it was found that the regions of interplanar defects have a short focal depth coming into and out of focus within a couple of angstroms. These results reveal the presence of interplanar defects aligning into columns after which they annihilate forming more extended graphitic structures. 

\begin{figure}[t!]
\centering
\includegraphics[width=0.9\linewidth]{Results2b.png}
\caption{\textbf{Observing graphitisation with electron microscopy}. Left are phase contrast transmission electron microscopy images and right are image processed skeletonised fringes. a) After a 10 seconds pulse at 2400~$^{\circ}$C\ a zig-zag texture of small fringes is found. Many interplanar defects appear. b) For 60 seconds of pulsing the interplanar defects have aligned into columns. c) These interplanar defects begin to annihilate forming extended fringes after 200~s. }
\label{fig:HRTEM}
\end{figure}

To consider what defects could be present, annealed molecular dynamics simulations were performed. 
Figure~\ref{fig:formation} a) shows the initial configuration for the simulations with aligned aromatic coronene and benzene fragments to mimic the aligned mesophase formed during carbonisation of graphitising carbons.
Molecular dynamics simulations were performed using the EDIP forcefield capable of accurately describing the bond forming and breaking in carbon materials as demonstrated in disordered carbons~\cite{Martin2019d,DeTomas2016}.
An elevated temperature of 3200~K was used to ensure dynamics on the simulation timescales of picoseconds.
Figure~\ref{fig:formation} b) shows the structure after the first ten picoseconds. A variety of non-hexagonal rings are formed, however, no stacking order was found. 
After 100~ps Figure~\ref{fig:formation} c) shows stacking and small regions with screw-like structures. At this stage extended hexagonal regions were also formed. 
By 1000~ps the main interplanar defect seen are screw dislocations with alignment of screw along the entire length of the box (Fig.~\ref{fig:formation} d)).
These simulations reveal that screw dislocations can form from an ordered phase of carbon resembling the mesophase. 

\begin{figure*}[t]
\centering
\includegraphics[width=0.9\linewidth]{ScrewFormation.png}
\caption{\textbf{Formation and imaging of screw defects.} a) shows the initial configuration of aligned aromatic regions. b) to d) shows snapshots of the structure from two orientations after the molecular dynamics simulations at 3200~K where the non-hexagonal rings are coloured. The graphenic regions of hexagonal rings are found to extend as the structure is annealed and screws form inbetween the layers c) followed by alignment d). e) shows a molecular model of a screw dislocation in graphite. f) simulated HRTEM images are shown infront, at focus and behind the dislocation core demonstrating the narrow range of focal depth. g) shows a similar defect experimentally imaged in the 2400~$^{\circ}$C for 200 s sample with HRTEM.}
\label{fig:formation}
\end{figure*}

The high resolution TEM image of a screw dislocation in ideal AA stacked graphite can be simulated and compared with the experimental fringes. 
Figure~\ref{fig:formation} e) shows a molecular model of the cylindrical core of a screw dislocation in AA graphite. The ramped structure connecting the layers together can be seen.
Figure~\ref{fig:formation} f) shows the simulated HRTEM images at focus as well as infront and behind. At focus the interdigited fringes offset by c/2 can be seen. Compared with the experimental HRTEM image for the 200s at 2400$^{\circ}$C sample (Figure~\ref{fig:formation} g) a close resemblance can be seen.
By defocusing by only 0.2~nm the interdigited structure disappears and a ramped structure is seen either to the left or to the right depending on the chirality of the screw. This short depth of field also match what is seen experimentally and eliminates attribution to other extended defects such as prismatic edge dislocations (or reconstructions of them) that can also provide the fringes offset by half a layer but would have a larger depth of field in the microscope or would shift predictably with focal depth.
The Supplementary information shows many screws in a sample revealing a wavy texture when the  interscrew distances are greater than $\sim$2~nm. This matches one of the last textures seen as graphite forms a series of wavy fringes.
The multislice HRTEM simulations of screw dislocations match the experimentally seen interplanar defects, depth of focus and wavy texture~\cite{Oberlin1984} providing a reasonable assignment for the defect involved in graphitisation at high temperatures.

To understand how screw dislocations can be removed further molecular dynamics simulations were performed at an increased heating rate. 
Similar simulations were performed as previously described, however, a slightly higher temperature was used of 3500~K enabling the screw glide to occur on picosecond timescales.
Figure~\ref{fig:annihilation} a) to c) shows a slice of the molecular structure with screw dislocations highlighted and with an arrow in the direction of the Burger's vector. At 739 picoseconds the two screws in the top left form a dislocation loop with two prismatic edge dislocations in the plane of the graphite layers. 
Figure~\ref{fig:annihilation} b) shows the glide of the topmost screws happening before 994 picoseconds reducing the length of the screw by one layer.
Figure~\ref{fig:annihilation} c) shows the structure after the pair annihilation of the two topmost screws dislocation. 
The removal of these defects is identical to that previously studied computationally for the ramp or mezzanine defect ~\cite{Trevethan2013,Vuong2017a,Vuong2017b}. This was found to be annihilated with a barrier of 3.24~eV in AA stacked graphite~\cite{Martinez2007}. This is a similar energy found for barrier for screw dislocation gliding in an similar system of 2.8~eV per layer. However, it should be noted in the case of the mezzanine defect only six carbon atoms are displaced so this energy is perhaps too low compared with an extended screw dislocation undergoing pair annihilation.
Using an Arrhenius argument the temperatures used can be scaled to experimental timescales (see Supplementary Information). This provides temperatures close to the critical temperature seen if 30~s is considered the time scale. Providing a barrier for graphitisation of 8~eV close to that measured in this study. 
These simulations reveal that screw dislocations can glide and when two of the opposite sense meet can annihilate providing insights into the kinetics and barriers involved in graphitisation.

\begin{figure}[t]
\centering
\includegraphics[width=\linewidth]{ScrewAnnihilation2.png}
\caption{\textbf{Screw annihilation and graphenic structure enlargement.} a) to c) shows three snapshots of slices of the molecular structure during molecular dynamics simulations performed at 3500~K. Between a) and b) a glide is shown where two screws shrink by a layer. Then the two screw dislocations are found to glide again and annihilate with the prismatic edge dislocations leaving the upper layers without interplanar defects. d) to f) shows the graphenic crystallites coloured according to their sizes with histograms shown in g) to i). The total area of aligned hexagonal regions increases after the glide and pair annihilation with the number of larger crystallites increasing accordingly. Thus showing the screw anniliation is connected to graphenic structure enlargement.}
\label{fig:annihilation}
\end{figure}

How are the graphenic crystallites impacted by the annihilation of screw dislocations?
These crystallites or $L_{a\{110\}}$, can be approximately tracked in the simulations by considering the hexagonal rings that are connected and aligned to within 3$^{\circ}$. Figure~\ref{fig:annihilation} d) to f) shows the  crystallites colour coded according to their size as well as the histogram of size plotted g) to i).
At the beginning of the time series a significant fraction of the structure are between 0 and 24~\AA\ in size.
After the top layers are freed from a screw defect the total area is increased from 83--94~nm\textsuperscript{2} and the regions of extended hexagonal regions grow up to 40~\AA.
These screw dislocations appear to pin the in-plane growth in a similar manner to how dislocation can inhibit growth of grain boundaries in metals \textcolor{red}{ref Irene\? Nigel?}.
These simulations therefore provide evidence that the annihilation of screw dislocations allow for the growth of the $L_{a\{110\}}$ thus connecting this defect with the crystallographic ordering seen in X-ray diffraction. 

Due to the limited size of the simulations they could no capture other features of the screw formation but some insights can be gained from comparison with other studies.
Due to the periodic approximation used, screws can only grow to the extent of the simulation box and the alignment of multiple columns was not observed. 
This could suggest that the annealing of a particular density of interplanar defects gives rise to ordered columns a set distance apart, which could perhaps be inferred from the size of the aromatic regions in the precursor material during carbonisation.
The small size also means that the long range dispersion forces would not have contributed significantly.
Screw defects are known to provide a small misalignment angle between layers due to the small negative curvature present in the screw. This provides an increased $d_{\{002\}}$ distance and could explain the turbostratic ordering seen. 
These misalignment angles could have interesting electronic properties with magic angle graphene having superconducting properties at low temperature~\cite{Cao2018}. Some preliminary experimental work looking at Moir\'{e} patterns seen along the basal plane suggest these misalignments are present~\cite{Boi2021}. Screw formation and controlled annihilation may provide a path for creating these regions in partially graphitised materials. However, there are other ways that lattice mismatch can occur such as in multiwalled nanotubes, carbon fullerene onions or stacked nanocones. So these conclusions only apply for highly ordered graphitising carbons.  

What about the ABA stacking structure?
As the screws are removed and larger regions of $L_{a\{110\}}$ are formed dispersion will provide a considerable torque between hexagonal regions to pull the layers into ideal ABA stacking alignment. But we hypothesis this can only occur after the majority of screw dislocations are annihilated.
This has previously been shown by the strong correlation between $L_{a\{110\}}$ and $d_{\{002\}}$. In the current work these two parameters are time dependent while $L_{c\{002\}}$ is not. Some interesting work on XRD peak shape has suggested a progressive model of alignment where pairs of layers begin to align before complete graphitisation~\cite{Puech2019}. This would be interesting to explore in the light of the screw annihilation mechanism presented.

These findings also provide possible routes for improving the production of synthetic and discovery of natural graphite. 
In particular previous simulations of disordered carbons~\cite{Martin2019d} provide a comparison to highlight what is different with graphitising carbons. Both graphitising and non-graphitising materials possess non-hexagonal rings as well as screw defects. 
But in non-graphitising carbons bent wavy structures still persist after screw annihilation, which have long been seen experimentally~\cite{Heidenreich1968}.
Our simulations suggest these could be due to the presence of disclinations as we found an excess of negative curvature in all disordered carbons simulated~\cite{Martin2019d} but not in the graphitising models shown. These could provide a possible non-annealable topological defect that cannot be removed via screw dislocation pair annihilation. 
Therefore our first suggestion for improving graphtisation is to form a well ordered mesophase so that no bent regions with disclinations can form. 

Mechanical approaches have been extensively studies to align the structure during carbonisation. For example elongational shear in carbon fibre manufacture, stress to form highly oriented pyrolytic graphite~\cite{Blackman1962} or shear in diamond anvils~\cite{Wong2019}. These all appear to encourage large regions of local molecular orientation. This may also explain the success of using templating approaches to form graphite from non-graphitising materials by enforcing a 2D orientation~\cite{Nishihara2012}. While the mesophase alignment is critical it is interesting to consider how mechanical shear, twist or stress could lower the barrier to screw dislocation annihilation providing potential new mechanical routes to graphite. It is also interesting to note that in pyrolytic graphite an even number of left handed and right handed screws have been found~\cite{Hennig1965}, whereas in natural graphite many superscrews can be seen to form~\cite{Rakovan2002}. This suggests that natural graphite with very large crystals forms from a spiral growth mechanism which has yet to be synthetically replicated.

Finally, chemicals can be used to catalyse graphitisation.
Intercalating metals have been extensively employed~\cite{Oya1979} and could provide strain in the interplanar region that could aid in screw dislocation gliding.
Non-carbon atoms such as silicon that more readily form sp$^3$ bonds could also lower the barrier for screw dislocation migration. It is interesting to consdier the Acheson process which uses SiC as a precursor that produces industrial scale graphite at significantly lower temperatures of 1700--2500~$^{\circ}$C.
These results could also suggest that using graphitic crystallites as a temperature probe in geology could also be challenging if impurity in the material could accelerate graphitisation at a lower temperature. It is also unclear how the addition of shear in geological rock would further modify the graphitisation process as discussed previously~\cite{Bonijoly1982}. 
Chemical or mechanical approaches that specifically target screw dislocation migration could therefore provide new routes for making synthetic graphite and predicting where natural graphite forms underground.

\section*{Methods}

\subsection*{Preparation of carbonised and graphitised samples}
Polyvinyl chloride was heated in a STF 1200 tube furnace over four hours to a carbonisation temperature of 1000$^\circ$C. The temperature was held for a further hour before passively cooling to room temperature. Argon, flowing at a constant 1.5-2~L/min maintained an inert atmosphere. The carbonised sample was then milled in a mortar and pestle before further heating.

Temperatures above 1000$^\circ$C were achieved using a customised furnace built by GBC scientific equipment. The system separates the joule heated graphite tube of an atomic absorption spectrometer from the spectrometer, leaving a bench top unit capable of heating to 3000$^\circ$C in a 1--2 seconds. Elevated temperatures can be maintained for several tens of seconds, whilst an inert cooling gas, typically argon, is followed through the tube to prevent combustion. Operation in a pulsed mode extends time at elevated temperature, a short pause between pulses allows the cooling system to manage the thermal load. The temperature profile for heating pulses is as follows: room temperature~-~100$^\circ$C in 5~seconds, ramping to desired temperature at 200$^\circ$~per~second, holding desired temperature for a further 10~seconds before cooling to room temperature over a few seconds. The graphite tube can house samples on the tens of milligram scale.

\subsection*{Characterisation of partially graphitised material}
X-Ray Diffraction (XRD) is performed in a Bruker D8 Advance Diffractometer with Bragg-Brentano geometry and a Cu K-$\alpha$ source. XRD specimens are prepared by placing a powder sample onto a low signal silicon wafer holder. Patterns are collected over a 2$\theta$ range of 10-90$^\circ$ with a dwell time of 1.5~seconds and step size of 0.3$^\circ$. Further information regarding the device configuration can be found in Ref. \cite{Putman:2018}. Transmission electron microscopy samples are prepared by grinding with ethanol in an agate mortar, the suspension is sonicated before deposition on 300 mesh lacey carbon grids. A JEOL F200 TEM fitted with a  cryogenic electron source is used to analyse samples nanostructure. With an accelerating voltage of 200~KeV the cold field emission gun produces a coherent electron beam which allows images to be collected with a low surface current. Specific care is taken to minimise beam damage by limiting sample time under the beam and using low beam intensities. Skeletonisation was performed based on the previously developed methodology from Botero \textit{et al.} but re-implemented in python with skimage with     the thresholding approach of Zack \textit{et al.}~\cite{Zack1977} replacing the Otsu method provided more interplanar features.

\subsection*{Simulations}
Initial configurations were prepared by packing coronene fragments of the same orientation into a XX~nm size cube using the software packmol~\cite{Martinez2009}, benzene fragments were then added until the density achieved reached 2~g/cc. The carbon environment dependent interatomic potential (EDIP) was used to describe the energy of the carbon network~\cite{Marks2000}, which has been shown to accurately describe the bonding in sp\textsuperscript{2} to sp\textsuperscript{3} carbons~\cite{DeTomas2016}. The structures were minimised and molecular dynamics simulations performed using the software LAMMPS ~\cite{Plimpton1995}. The Bussi thermostat maintained the temperature of the simulation~\cite{Bussi2007}. The results were processed in polypy to extract the rings using the Flanbau ring algorithm~\cite{Franzblau1991} and custom scripts extracted graphenic regions. Finally all of the visualisations were done in the software VMD~\cite{Humphrey1996}. Atomsk~\cite{Hirel2015} was used to prepare screw dislocation defects in AA graphite. Multislice software compuTEM~\cite{Kirkland2010} was then used to simulate the HRTEM images of the dislocations using aberrations matching those of the instrument.

\bibliography{references.bib}

\noindent\textbf{Acknowledgements}\\
This project is supported by the Forrest Research Foundation.\\

\noindent\textbf{Author contributions}\\
K.K., F.S, M.C., A.D. and L.G. carried out the experimental work. J.W.M. and A.M. carried out the quantum calculations. J.W.M. and L.P. carried out the classical molecular dynamics and quantum molecular dynamics calculations and analysis. J.W.M.,  authors discussed and co-wrote the manuscript. \\

\noindent\textbf{Competing interests}\\
The authors declare no competing interests. \\

\noindent\textbf{Supplementary information}
This file contains Supplementary Figures S1--S12 and Supplementary Movies 1--5

\setcounter{figure}{0}
        \renewcommand{\thefigure}{E\arabic{figure}}